\documentclass{article}

\PassOptionsToPackage{numbers, compress}{natbib}

    \usepackage[final]{neurips_2023}

\usepackage[utf8]{inputenc} 
\usepackage[T1]{fontenc}    
\usepackage[colorlinks]{hyperref}       
\usepackage{url}            
\usepackage{booktabs}       
\usepackage{amsfonts}       
\usepackage{nicefrac}       
\usepackage{microtype}      
\usepackage[dvipsnames]{xcolor}         
\usepackage{amsmath}
\usepackage[english]{babel}
\usepackage[pdftex]{graphicx}
\usepackage{bm}
\usepackage{booktabs}
\usepackage{svg}

\title{Dynamic Neural Fields for Learning Atlases of 4D Fetal MRI Time-series}

\author{%
Zeen Chi$^{1,2}$\thanks{Equal contribution. Work done while visiting MIT CSAIL.} \quad Zhongxiao Cong$^{1,2}$\footnotemark[1] \quad \textbf{Clinton J. Wang}$^2$ \quad \textbf{Yingcheng Liu}$^2$ \\ \textbf{Esra Abaci Turk}$^{3,4}$ \quad \textbf{P. Ellen Grant}$^{3,4}$ \quad \textbf{S. Mazdak Abulnaga}$^{2,4,5}$ \\ \textbf{Polina Golland}$^2$ \quad \textbf{Neel Dey}$^2$ \\
$^1$School of Information Science and Technology, ShanghaiTech University \quad $^2$MIT CSAIL \\  $^3$Fetal-Neonatal Neuroimaging \& Developmental Science Center, Boston Children's Hospital \\ $^4$ Harvard Medical School \quad $^5$Massachusetts General Hospital 
\\
\texttt{\{chize,congzhx\}@shanghaitech.edu.cn}\quad\texttt{\{dey,abulnaga,liuyc\}@mit.edu}\\
\texttt{\{clinton,polina\}@csail.mit.edu}\\
\texttt{\{esra.abaciturk,ellen.grant\}@childrens.harvard.edu}
}

\begin{document}

\maketitle

\begin{abstract}


We present a method for fast biomedical image atlas construction using neural fields. Atlases are key to biomedical image analysis tasks, yet conventional and deep network estimation methods remain time-intensive. In this preliminary work, we frame subject-specific atlas building as learning a neural field of deformable spatiotemporal observations. We apply our method to learning subject-specific atlases and motion stabilization of dynamic BOLD MRI time-series of fetuses \textit{in utero}. Our method yields high-quality atlases of fetal BOLD time-series with $\sim$5-7$\times$ faster convergence compared to existing work. 
While our method slightly underperforms well-tuned baselines in terms of anatomical overlap, it estimates templates significantly faster, thus 
enabling rapid processing and stabilization of large databases of 4D dynamic MRI acquisitions. Code is available at \url{https://github.com/Kidrauh/neural-atlasing}.

\end{abstract}

\section{Introduction}
Given biomedical image observations, constructing image atlases enables morphometric analyses and registration to a common coordinate system. 
Current conventional \cite{avants2009advanced,kuklisova2011dynamic,makropoulos2016regional,rueckert1999nonrigid,schuh2015construction,serag2012multi} and deep learning methods \cite{chen2021construction,dalca2019learning,dey2021generative,zhang2016consistent,zhao2021learning} for atlas building yield high-quality atlases with accurate registration at the cost of significant computation time. These computational costs compound further when given \textit{subject-specific} image time-series (e.g., longitudinal repeats) where a new atlas must be constructed for each subject to enable motion stabilization and standardized analyses.

In the context of fetal image analysis, \textit{in-utero} BOLD MRI time series can track changes in fetal and placental oxygenation under induced maternal hyperoxia to identify dysfunction and monitor fetal and maternal well-being~\cite{aimot2013vivo,luo2015human,schopf2012watching,sorensen2013changes}. However, the inter-timepoint motion caused by fetal movement and maternal breathing necessitates nonlinear registration of the time series to a common coordinate system for each individual subject to stabilize motion prior to any analysis. To that end, this work presents a method for fast subject-specific spatiotemporal atlasing.



We formulate atlas estimation as the learning of compactly-parameterized dynamic neural fields~\cite{park2021nerfies,park2021hypernerf,pumarola2021d,song2022nerfplayer} to represent both the atlas and image-to-atlas deformations. 
Using our proposed neural representation and training strategy, we rapidly construct high-fidelity
subject-specific atlases and stabilize the motion present in BOLD MR images of fetuses \textit{in utero} to enable improved analyses of key BOLD time series-based fetal and maternal biomarkers~\cite{sorensen2013changes}.


\begin{figure}[!ht]
    \centering
    \includegraphics[width=\textwidth]{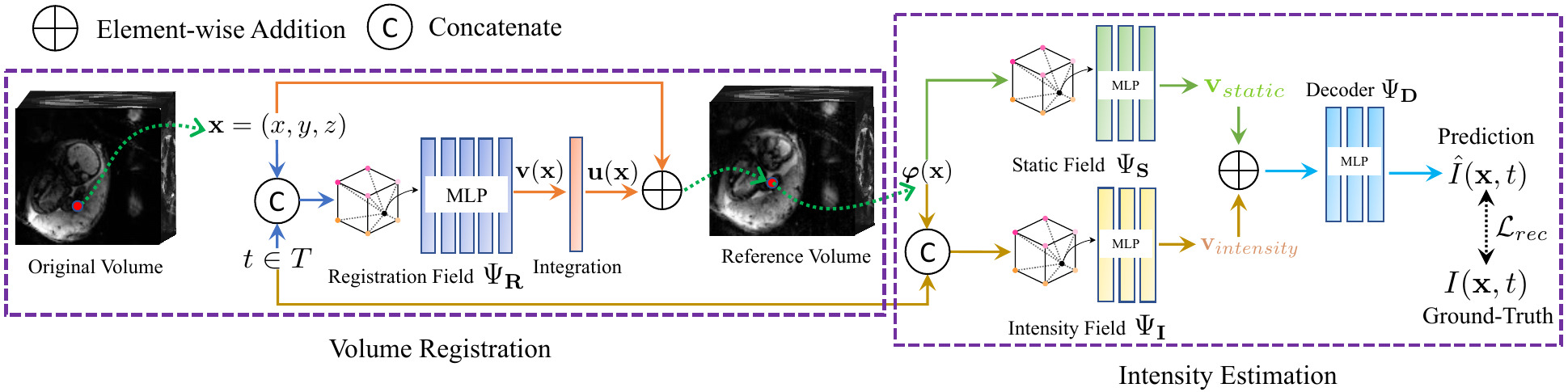}
    \vspace{-5pt}
    \caption{\textbf{Architecture.} Our method constructs neural fields for volume registration and intensity estimation, which warp observations to an atlas space and learn the atlas parameters, respectively. 
}
    \label{fig:pipeline}
\end{figure}


\section{Methods}
\noindent\textbf{Learning Neural Fields.} 
Fig.~\ref{fig:pipeline} presents our method consisting of networks for image-to-atlas deformation and atlas estimation. We use three neural fields, each parameterized as a multi-resolution hash encoding followed by a small MLP~\cite{muller2022instant} for efficient processing. We further use stationary velocity fields (SVF) to ensure diffeomorphic deformations~\cite{arsigny2006log,ashburner2007fast,modat2012parametric}.
The atlas is produced by an encoder-decoder where the encoder consists of time-invariant (\textit{static}) and time-variant (\textit{intensity}) functions that allow small changes in atlas appearance to account for subtle topological changes.

Given spatial $\mathbf{x}=(x,y,z)$ and temporal $t\in T$ coordinates, the registration field $\Psi_{\mathbf{R}}:\mathbb{R}^4\mapsto\mathbb{R}^3$ computes velocities $\mathbf{v(x)}$ which integrate to yield a diffeomorphic displacement field $\mathbf{u(x)}$ between an image at time $t$ and the atlas, such that the deformation between them is $\bm{\varphi}(\mathbf{x})=\mathbf{u(x)}+\mathbf{x}$. On warping the image coordinates into the atlas space, we query $\bm{\varphi}(\mathbf{x})$ from the static field $\Psi_{\mathbf{S}}:\mathbb{R}^3\mapsto\mathbb{R}^n$ to get the feature vector $\mathbf{v}_{static}\in\mathbb{R}^n$ encoding time-invariant latent atlas features. We then query $(\bm{\varphi}(\mathbf{x}),t)$ from an intensity field $\Psi_{\mathbf{I}}:\mathbb{R}^4\mapsto\mathbb{R}^n$ that yields $\mathbf{v}_{intensity}\in\mathbb{R}^n$ encoding the latent intensity differences between $\bm{\varphi}(\mathbf{x})$ in the atlas and $\mathbf{x}$ in the original image. An MLP $\Psi_{\mathbf{D}}:\mathbb{R}^n\mapsto\mathbb{R}$ then decodes the fused latent features and yields the estimated intensity $\hat{I}(\mathbf{x},t)$ of the original image. 


\begin{table}[!t]
    \centering
    \footnotesize
    \caption{\textbf{Quantitative results} of baseline comparisons (\textbf{top}) and ablations (\textbf{bottom}) studying registration performance (via local normalized cross-correlation and weighted dice), deformation qualities (via deformation magnitude, avg. Jacobian determinant, and folding ratio), and runtimes.}
    \begin{tabular}{c@{\hspace{0.15cm}}c@{\hspace{0.15cm}}c@{\hspace{0.15cm}}c@{\hspace{0.15cm}}c@{\hspace{0.15cm}}c@{\hspace{0.15cm}}c}
    \toprule
        & \textbf{LNCC} ($\uparrow$) & \textbf{Wt. Dice} ($\uparrow$) & $\lVert\mathbf{u(x)}\rVert_2$  ($\downarrow$) & $|J_{\bm{\varphi}}|$ & \textbf{\% folds} ($\downarrow$) & \textbf{Runtime} ($\downarrow$) \\
        \midrule
        Unaligned & 0.392(0.073) & 0.80(0.05) & - & - & - & - \\
        SyGN~\cite{avants2010optimal} & 0.528(0.075) & \textbf{0.91}(0.02) & 0.0227(0.0035) & 1.000(0.000) & \textbf{0} & 12hrs / 96-core CPU \\
        AtlasMorph~\cite{dalca2019learning} & 0.531(0.079) & 0.90(0.02) & \textbf{0.0083}(0.0014) & 1.004(0.003) & \textbf{0} & 16hrs / A6000 GPU \\
        Ours & \textbf{0.579}(0.081) & 0.88(0.02) & 0.0183(0.0067) & 1.004(0.013) & 0.01(0.01) & \textbf{2.2}hrs / A6000 GPU \\ \midrule
        (- SVF) & 0.503(0.081) & 0.85(0.04) & 0.0096(0.0021) & 1.006(0.010) & 0.04(0.02) & 1.1hrs / A6000 GPU \\
        (- Divergence) & 0.579(0.078) & 0.87(0.02) & 0.0200(0.0063) & 1.013(0.012) & 0.06(0.04) & 1.5hrs / A6000 GPU \\
        (- Intensity field) & 0.578(0.083) & 0.88(0.02) & 0.0209(0.0086) & 1.000(0.018) & 0.01(0.01) & 2.2hrs / A6000 GPU \\
    \bottomrule
    \end{tabular}
    \label{tab:results}
\end{table}

\noindent\textbf{Losses.} 
We use the $L_1$ reconstruction objective 
$\mathcal{L}_{rec}=\frac{1}{|\Omega|}\sum_{\mathbf{x}\in\Omega}|I(\mathbf{x},t)-\hat{I}(\mathbf{x},t)|$ where $\Omega$ is the spatial coordinates and $I$ and $\hat{I}$ are ground truth and estimated intensities of the image, respectively. To encourage smooth, locally-rigid, and central deformations, we develop the regularizer $\mathcal{L}_{def}=\lambda_1\frac{1}{|\Omega|}\sum_{\mathbf{x}\in\Omega}\lVert\mathbf{u(x)}\rVert_2+\lambda_2\mathcal{L}_{div}+\lambda_3\lVert\mathbf{\bar{u}(x)}\rVert_2^2$, where $\mathbf{\bar{u}(x)}$ is the moving average of displacement vectors~\cite{dalca2019learning} and $\mathcal{L}_{div}=\frac{1}{|\Omega|}\sum_{\mathbf{x}\in\Omega}|\mathrm{div}(\mathbf{u(x)})|^2$ is the divergence loss \cite{tretschk2021non} that encourages locally-rigid deformations which are essential to properly model fetal motion.
To reduce folds in the computed deformations, we use the negative Jacobian loss $\mathcal{L}_{jac}$ \cite{mok2020large}, which reduces the number of negative elements in the determinant of the Jacobian of the deformation. For intensity estimation, we use $L_1$ regularization $\mathcal{L}_{int}$ on $\mathbf{v}_{intensity}$ to limit temporal appearance changes, and use total variation regularization $\mathcal{L}_{tv}=\mathrm{tv}(\mathbf{v}_{static})+\mathrm{tv}(\mathbf{v}_{intensity})$ on $\mathbf{v}_{static}$ and $\mathbf{v}_{intensity}$ to encourage 
piecewise-constant and sharp-edged atlases both spatially and temporally. Our overall objective is $\mathcal{L}(F)=\mathcal{L}_{rec}+\mathcal{L}_{def}+\lambda_{jac}\mathcal{L}_{jac}+\lambda_{int}\mathcal{L}_{int}+\lambda_{tv}\mathcal{L}_{tv}$ where 
$\lambda_1=10^{-3}, \lambda_2=5 \times 10^{-4}, \lambda_3=0.1, \lambda_{jac}=1, \lambda_{int}=0.05$, and $\lambda_{tv}=0.1$, chosen via grid search on two validation subjects.

\noindent\textbf{Atlas Inference.}
To construct the final atlas (the single time-invariant template) representing the entire time-series, we directly query $(\mathbf{x}, t)$ from the trained atlas encoder-decoder network (Fig.~\ref{fig:pipeline} right, intensity estimation). We first calculate the static feature vector $\mathbf{v}_{static}$ and the intensity feature vectors $\mathbf{v}_{intensity}$ at each time step $t$ and then decode $\mathbf{v}_{static}+\frac{1}{T}\sum_{t=1}^T\mathbf{v}_{intensity}$ using $\Psi_{\mathbf{D}}$.


\begin{figure}
    \centering
    \includegraphics[width=\textwidth]{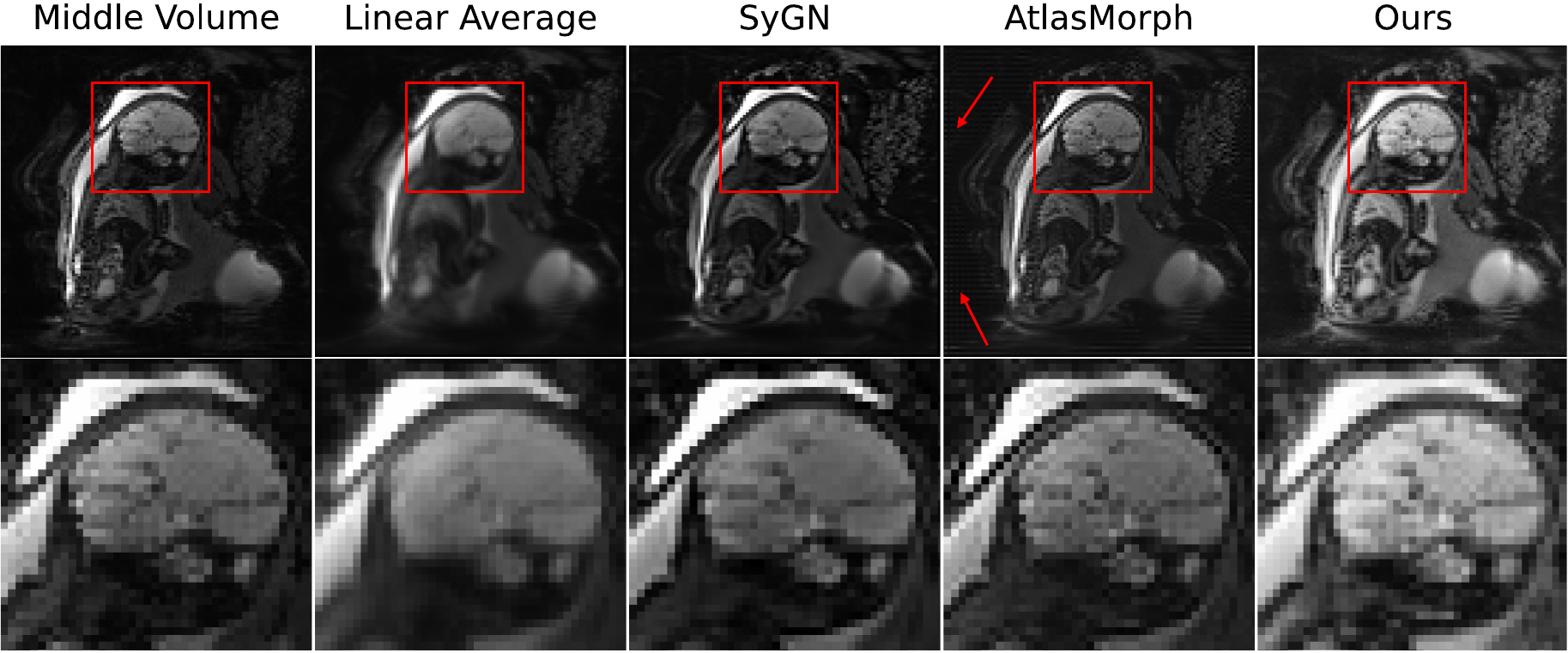}
    \caption{Given an arbitrarily chosen subject, we illustrate the mid-timepoint of the time-series, the temporal linear average, and fetal atlases produced by SyGN~\cite{avants2010optimal}, AtlasMorph~\cite{dalca2019learning}, and our method. Atlasmorph creates undesirable checkerboard artifacts (indicated by red arrows).}
    \label{fig:atlases}
\end{figure}

\section{Experiments}
\noindent\textbf{Data and Baselines.} We use 11 dynamic BOLD MRI time-series of \textit{in utero} fetal subjects (2 for tuning hyperparameters and modeling decisions and 9 for held-out testing) with a time-series length of 78 to 146 time points per subject.
Due to fetal motion and maternal breathing, there is a need for registration of all images to a common unbiased subject-specific representation \cite{abulnaga2022automatic}. Each image is resampled to $112\times 112\times 80$ at $3 mm^{3}$ isotropic resolution. As we use an intensity-based reconstruction loss, we use adaptive histogram equalization~\cite{pizer1987adaptive} for inputs to our model to balance contributions from bright and dark BOLD regions such as the amniotic fluid and fetal body, respectively. We use SyGN~\cite{avants2010optimal} and AtlasMorph~\cite{dalca2019learning} as representative conventional and deep network baselines, with local normalized cross-correlation (LNCC)~\cite{avants2008symmetric} as a registration loss which is locally-adaptive and intensity scale-invariant by design. AtlasMorph and our method are trained on a single NVIDIA RTX A6000 GPU and SyGN is optimized on a server CPU using 96 hyperthreaded cores.

\noindent\textbf{Evaluation.} Atlas building evaluation is subtle and involves trade-offs between registration accuracy, deformation quality, and runtime~\cite{dey2021generative}. To measure performance, we follow~\cite{ding2022aladdin} and randomly select 50 MRI pairs for each subject and compose image-to-atlas and atlas-to-image warps to calculate LNCC and multi-region Dice coefficients \cite{dice1945measures}. Our segmentation labels correspond to the placenta, amniotic fluid, fetal body, fetal brain, and fetal eyes and are generated by an in-house segmentation network. To assess deformation quality, we calculate the average displacement $L_2$ norm between the atlas and images with a lower value indicating improved template centrality, the mean determinant of the Jacobian matrix $J_{\bm{\varphi}}(p)$ w.r.t. the input voxel $p$, and the ratio of deformation folds. 

\noindent\textbf{Results.} Table~\ref{tab:results} reports LNCC and the weighted average Dice scores and deformation statistics comparisons between the baselines and our model. All methods produce invertible deformations. The proposed model achieves best-in-class LNCC but lags behind slightly in terms of Dice score (i.e., anatomical overlap). In terms of runtime, our proposed model converges $5.5-7.4\times$ faster than baselines yielding high-fidelity templates (see Fig.~\ref{fig:atlases}) with smooth and invertible deformations. However, if the tuned baselines are optimized to convergence, they currently yield improved anatomical overlap. Ablations removing the SVF formulation, the divergence loss, and $\Psi_{\mathbf{I}}$ all worsen performance.

\section{Conclusions and Future Directions}
We demonstrate that dynamic neural fields learn atlases of 4D fetal BOLD MRI time-series significantly faster than current methods. These speed gains are especially relevant to subject-specific atlas building of large collections of subjects imaged using 4D dynamic MRI. Currently, our preliminary work finds that well-tuned baselines optimized for longer still achieve better registration overlap in terms of Dice. This performance gap points to several future directions: 
(1) Fetal BOLD MRI time series are temporally sampled at only $\sim$0.28 frames per second (FPS) as compared to conventional video (24+ FPS) for which existing work on dynamic neural fields was developed. This gives rise to large, erratic motion between consecutive timepoints, and may require modification to existing positional encoding functions which assume temporal smoothness. (2) High-performing mono-modal biomedical image registration frameworks typically use LNCC~\cite{avants2011reproducible} as a registration loss. However, due to the scale and shift-invariant formulation of LNCC, neural regression networks trained with LNCC require significant regularization to guide them towards non-degenerate solutions, which we find 
can introduce significant artifacts in the estimated atlas.
Future work may seek to mitigate this tradeoff by constraining the optimization space of the network or using data-driven priors.

\section{Acknowledgements}
We gratefully acknowledge funding from NIH NIBIB 5R01EB032708, NIH NICHD R01HD100009, NIH NIA 5R01AG064027, and NIH NIA 5R01AG070988. We thank all the participants for providing the data in the study.


{
\small

\bibliography{neurips_2023}
}


\end{document}